\documentclass[twocolumn,superscriptaddress,nofootinbib]{revtex4}

\usepackage{amssymb}
\usepackage{amsmath}
\usepackage{multirow}
\usepackage{mathrsfs}
\usepackage{graphicx}
\usepackage{color}
\usepackage{dcolumn}
\usepackage{bm}
\newfont{\Fr}{eufm10}
\usepackage{graphics}
\usepackage{epsfig}
\usepackage[english]{babel}
\usepackage{verbatim}
\usepackage{amsfonts}
\usepackage[latin1]{inputenc}
\usepackage{hyperref}

\newcommand{\tc}{\eta_k^c}
\newcommand{\beq}{\begin{equation}}
\newcommand{\eeq}{\end{equation}}
\newcommand{\barr}{\begin{eqnarray}}
\newcommand{\earr}{\end{eqnarray}}
\newcommand{\nk}{\vec{k}}
\newcommand{\dphi}{\delta \phi}
\newcommand{\x}{\vec{x}}

\newcommand{\bra}{\langle}
\newcommand{\ket}{\rangle}
\newcommand{\mH}{\mathcal{H}}

\newcommand{\hv}{\hat{v}}
\newcommand{\hp}{\hat{\pi}}
\newcommand{\h}{\mathcal{H}}

\newcommand{\nx}{\vec{x}}
\newcommand{\hb}{\hat{\beta}}

\newcommand{\hh}{\hat{h}}

\begin{document}

\title{Inflationary gravitational waves in collapse scheme models}
\author{Mauro Mariani}
\email{mariani@carina.fcaglp.unlp.edu.ar} \affiliation{Facultad de
Ciencias Astron\'{o}micas y Geof\'\i sicas, Universidad Nacional de La
Plata, Paseo del Bosque S/N, 1900 La Plata, Argentina}
\author{Gabriel R. Bengochea}
\email{gabriel@iafe.uba.ar} \affiliation{Instituto de Astronom\'\i
a y F\'\i sica del Espacio (IAFE), UBA-CONICET, CC 67, Suc. 28,
1428 Buenos Aires, Argentina}
\author{Gabriel Le\'{o}n}
\email{gleon@df.uba.ar} \affiliation{Departamento de
F\'\i sica, Facultad de Ciencias Exactas y Naturales, Universidad
de Buenos Aires, Ciudad Universitaria - Pab.I, 1428 Buenos Aires,
Argentina}

\begin{abstract}

The inflationary paradigm is an important cornerstone of the
concordance cosmological model. However, standard inflation
cannot fully address the transition from an early
homogeneous and isotropic stage, to another one lacking such
symmetries corresponding to our present universe. In previous
works, a self-induced collapse of the wave function has been
suggested as the missing ingredient of inflation. Most of the
analysis regarding the collapse hypothesis has been solely focused
on the characteristics of the spectrum associated to scalar
perturbations, and within a semiclassical gravity framework. In
this Letter, working in terms of a joint metric-matter
quantization for inflation, we calculate, for the first time, the
tensor power spectrum and the tensor-to-scalar ratio corresponding
to the amplitude of primordial gravitational waves resulting from
considering a generic self-induced collapse.

\end{abstract}

\pacs{Valid PACS appear here}


\maketitle

\section{Introduction}

\label{intro}

The vast majority of the cosmological community considers
the inflationary paradigm on a stronger footing than ever given
the agreement between its predictions and the latest
observations (e.g. WMAP9 \cite{wmap9}, Planck \cite{planck1}). In
particular, last year a claim by BICEP2 Collaboration regarding the
detection of primordial tensor modes \cite{bicep}, in spite of the
subsequent controversy of their results \cite{flauger, seljak, planck2},
has made some cosmologists think that this important prediction from the
traditional inflation model will be confirmed in the foreseeable
future, which in turn will reassert the standing of the model.

According to the traditional inflationary paradigm, the early universe undergoes an
accelerated expansion (lasting at least some 70 e-folds or so), resulting in an
essentially flat, homogeneous and isotropic  space-time with an extreme dilution of
all unwanted relics. Note that the dynamics of the space-time is governed by Einstein
equations which are symmetry preserving, i.e. the symmetry being the homogeneity
and isotropy (H\&I). Another important aspect is that when considering the quantum
features of the scalar field (the inflaton) driving the expansion. This field, is assumed
to be in the vacuum state as a result of the same exponential expansion, and one finds also that it
contains ``fluctuations'' with the appropriate nearly-scale-invariant spectrum.
These vacuum fluctuations are considered responsible for all the structures we
observe in the actual universe, and in particular, the observed cosmic microwave
background (CMB) anisotropies.

One cannot deny the favorable matching between the model
predictions and observations; nevertheless, from the conceptual
point of view something is missing. Even if the inflaton contains quantum
uncertainties (or vacuum fluctuations), according to the Quantum Theory, the
physical state of the system is encoded in the quantum state. The vacuum state of the
quantum fields is H\&I, i.e. it is an eigen-state of the operators generating spatial
translations and rotations (see Appendix A of Ref. \cite{stat} for a proof). The fact
that a system contains quantum uncertainties does not necessarily implies that
it contains actual inhomogeneities and anisotropies, since the quantum state, which
characterize the physical state of the system, can still be perfectly H\&I. Additionally, the dynamics
of the quantum state is governed by Schr\"{o}dinger equation, which does not break
translational and rotational invariance. Consequently, the initial quantum state
cannot be evolved into a final state lacking such symmetries. Thus, there is an important
issue, namely: what is the precise mechanism by which the primordial perturbations are
born given that the equations governing the dynamics are symmetry preserving? In other
words, it is not clear how from an initial condition that is
H\&I (both in the background space-time and in
the quantum state that characterizes the quantum fields), and based
on a dynamics that supposedly preserves those symmetries, one ends
up with a non-homogeneous and non-isotropic state associated to
the late observed universe.

The above described situation is sometimes related to the issue of
the quantum-to-classical transition of the primordial quantum
fluctuations. And, although decoherence provides a partial
understanding of the issue \cite{kiefer,jmartin}, it does not
fully address the problem; mainly because decoherence does not
solve the quantum measurement problem, which appears in an
exacerbated manner in the case of the inflationary universe. We
invite the interested reader to consult, for instance, Refs. \cite{pss, shorts} where a more detailed analysis has
been made regarding the issues with decoherence and other
approaches to the problem at hand.

In order to account for the aforementioned problem, Sudarsky et
al. \cite{pss} proposed a self-induced collapse of the wave
function, i.e. a spontaneous change from the original quantum
state associated to the inflaton field into a new quantum state
lacking the symmetries of the initial state. Also, their approach relies on the
semiclassical gravity framework, in which matter is described by a Quantum Field
Theory and the space-time is always treated in a classical manner. The self-induced
collapse is considered as being the responsible of generating the primordial
perturbations. In
particular, by relying on Einstein semiclassical equations, the expectation value in the
post-collapse state of the quantum matter fields is related to the metric of the
space-time which is always classical. The result of the
evolution of the metric perturbations, born after the collapse,  is related to the actual
anisotropies and inhomogeneities observed in the CMB radiation.
Thus, in this proposal, after the collapse, the universe is described by
a space-time and a quantum state that are no longer H\&I.

On the other hand, it is evident that the collapse mechanism should be a physical
process independent of external entities, since in the early universe there is not a clear
notion of observers, measurement devices, environment, etc. It is worthwhile to comment
that models involving an objective dynamical reduction of the wave function (in
different contexts from cosmology) have been proposed in past years
\cite{diosi,penrose,GRW,pearle,pearle2,bassi}.
These models attempt to provide a solution to the so-called measurement problem of
Quantum Mechanics by eliminating from the theory the need of an external agent
responsible for localizing the wave function. It
is also interesting that these models give predictions that can be tested experimentally
and that are different from the standard Quantum Theory \cite{mohammed}. We will not deal
with all the conceptual framework concerning the self-induced
collapse and instead we will refer the interested reader to Refs.
\cite{stat,pss,shorts,alberto} for a more in depth
analysis.

Previous works, e.g. \cite{pss, adolfo2008, gabriel2010},
have analyzed the characteristics of the spectrum associated with
the scalar perturbations resulting from considering the
self-induced collapse hypothesis in different inflationary scenarios, e.g.
multiple collapses \cite{multiples}, correlation between the modes
caused by the collapse \cite{sigma}, collapse occurring during the
radiation dominated era \cite{susana2014}, and also in a non-inflationary model \cite{noninfl}. Moreover, in Ref.
\cite{susana2012} two quantum collapse schemes were tested with
recent data from the CMB, including the 7 year release of WMAP \cite{wmap7}
and the matter power spectrum measured using LRGs by
the Sloan Digital Sky Survey \cite{SDSS}. However, as we have mentioned, most previous
mentioned works have been based on the semiclassical gravity
approximation, which enables a quantum treatment of the matter
fields, while a classical description of gravitation is
maintained. In particular, the amplitude of primordial tensor
modes provided by the collapse hypothesis, within the
semiclassical gravity approximation, is exactly zero at
first-order in perturbation theory \cite{pss,gabriel2010}. At second-order, the
model prediction for the amplitude is too low that is
practically undetectable by any recent and future experiments \cite{secorder}.

On the other hand, last year an allegation concerning the detection of
primordial $B$-modes polarization of the CMB by BICEP2
Collaboration \cite{bicep} (notwithstanding the apparent tension with the
results provided by Planck mission and a strong evidence of
probable contamination by Galactic dust \cite{bicepplanck}), has made the revelation
of primordial gravity waves a real possibility. In the plausible
scenario of a confirmed detection of primordial $B$-mode
polarization, the framework of semiclassical gravity applied to
the inflationary universe faces several issues, nevertheless, one could
still implement the self-induced collapse hypothesis. One possible
option (and probably the simplest) is to apply the collapse
proposal directly within the standard analysis, in terms of a
quantum field jointly characterizing the inflaton and metric
perturbations, the so-called Mukhanov-Sasaki variable. In Ref. \cite{gabriel} a first step, regarding the implications of
considering the collapse of the wave function characterizing the
state of the quantum field associated to the Mukhanov-Sasaki
variable, was made. In particular, it was shown that the standard
shape of the spectrum associated to the scalar perturbations
becomes altered by introducing the collapse hypothesis.
Furthermore, in Refs. \cite{jmartin2,hinduesS} a particular objective
collapse model, called Continuous Spontaneous Localization (CSL)
collapse model \cite{GRW,pearle,pearle2}, was implemented resulting in interesting
modifications to the standard scalar power spectrum corresponding
to the Mukhanov-Sasaki variable field.

In this work, we will make a step further and obtain the spectrum
associated to the tensor modes within the framework of quantizing
both the matter and metric perturbations. We will show that, as in
the scalar case, the tensor power spectrum becomes modified by
introducing the collapse hypothesis. Additionally, we will obtain
the tensor-to-scalar ratio $r$ and show that it is of the same
order of magnitude as the one predicted by standard single-field
slow-roll inflation. Nevertheless, an interesting result is that
$r$ is independent of the collapse parameters. Thus, the precise
measurement of $r$ sets the energy scale of inflation (the same as
in the standard case), but cannot yield any significant information
concerning the collapse. Moreover, we will not consider a specific
collapse mechanism, but we will parameterize the
collapse generically through the expectation values of the field and its
conjugated momentum evaluated in the post-collapse state.
It is worthwhile to mention that, in Ref. \cite{hinduesT}, the CSL
collapse model was used to analyze the tensor modes in the same
context as the present work, in relation to the quantum treatment
of the fields. The authors conclude that accurate
measurements of $r$ and the tensor spectral index $n_T$ can help
to constraint such model parameters. However, their point of
view regarding the physical implications of the collapse is
different from ours. Specifically, in our picture if there is no
quantum collapse the quantum state of the field is homogeneous and isotropic and there
are no perturbations of the space-time, thus $r=0$. On the other
hand, within the model analyzed in \cite{hinduesT}, in the absence
of a quantum collapse one recovers the standard inflationary predictions concerning the
tensor and scalar power spectra. This is an important distinction with further
implications regarding the observational quantities, as it will be
shown in future work, but more importantly it constitutes a difference in the physical implication of the self-induced collapse.

The present Letter is organized as follows: in Section \ref{review} we review some basics about previous results regarding the power spectrum of scalar perturbations, in the framework of collapse scheme models and working in terms of a joint metric-matter
quantization for inflation; in Section \ref{tensor} we show our results for the power spectrum of tensor modes and the tensor-to-scalar ratio; and finally, in Section \ref{conclusions} we summarize our conclusions.

\section{Brief review of previous results}
\label{review}

In this Section, we will present a brief review of the results
obtained in Ref. \cite{gabriel}, where the self-induced
collapse hypothesis was added to the standard quantum treatment
characterizing the primordial perturbations, namely to the scalar
field associated to the Mukhanov-Sasaki variable. Specifically, we
will mention the problem with the standard picture, and then
we will motivate the addition of the self-induced collapse. Later, we will focus
on the power spectrum corresponding to the scalar perturbations
within our model. There is no original work in this Section, and
detailed analyses can be found in Refs. \cite{pss,shorts,stat,gabriel2010}.

We start characterizing the inflationary universe by Einstein
theory $G_{ab} = 8\pi G T_{ab}$ ($c=1$) along with the dynamics of the
matter fields corresponding to the inflaton. Also, we shall work
with the standard single-field slow-roll inflaton $\phi$.
Specifically, the background space-time is described by an approximately de
Sitter expansion. Thus, the scale factor, in conformal time $\eta$, is given by $a(\eta) \simeq
-1/H\eta$ with $H$ the Hubble parameter, approximately constant. On the other
hand, the matter sector is dominated by the inflaton, which is
``rolling slowly'' down the potential $V$; consequently, the
slow-roll parameter is defined $\epsilon \equiv 1-\mH'/\mH^2$. Here, a prime denotes partial derivative with respect to conformal time $\eta$,
and $\mH \equiv a'/a$ is the conformal expansion rate. Also, during slow-roll inflation $\epsilon \simeq M_P^2/2 (\partial_\phi V/V)^2$ where $M_P^2 \equiv (8\pi G)^{-1}$ is the
reduced Planck mass; additionally, we will work with the assumption that $\epsilon =$ constant.\footnote{As it is well known, assuming $\epsilon$ to be exactly constant
leads to a perfect scale-invariant power spectrum
(scalar and tensor). It is only by considering $\epsilon'\neq 0$, also known as a quasi-de Sitter
expansion, that one obtains a small dependence on $k$ in the
traditional power spectrum of the form $k^{n_s-1}$ with $n_s \neq 1 $. On the other hand, in this work we
are mainly interested in the amplitude and not in the shape of the
tensor spectrum. Therefore, we can perform all the analysis in an
the approximation that $\epsilon$ is exactly a constant, without loss of generality, and
finally obtain the amplitude of the corresponding spectra.}

We choose to work in the longitudinal gauge, and we assume no
anisotropic stress. So, the scalar perturbations of the metric are
represented, in comoving coordinates, by the following line element:
\beq ds^2 = a^2(\eta) [-(1+2\Psi) d \eta^2 + (1-2\Psi) \delta_{ij}
dx^i dx^j ]\eeq
with $\Psi(\eta,\x) \ll 1$. Decomposing the scalar field into an homogeneous and isotropic
part plus small perturbations $\phi(\x,\eta) = \phi_0 (\eta) +
\dphi (\x,\eta)$, one can construct the Mukhanov-Sasaki variable
\beq\label{defv} v \equiv a \left( \dphi + \frac{\phi_0'}{\mH}
\Psi \right). \eeq
 Einstein perturbed equations
at first-order $\delta G_{ab}= 8 \pi G \delta T_{ab}$, imply
\beq\label{EEs} \nabla^2 \Psi = - \sqrt{\frac{\epsilon}{2}}
\frac{H}{M_P} \left(v'-\frac{z'}{z} v \right) \eeq
where $z\equiv a\phi_0'/\mH$. Moreover, since we are assuming an approximately de Sitter
expansion, i.e. assuming $\epsilon' =0$ and slow-roll type of inflation, then $z'/z = a'/a$. It is
important to mention that in the longitudinal gauge, the field
$\Psi$ represents the curvature perturbation of the background and is related to the Mukhanov-Sasaki variable $v$ as in Eq. \eqref{EEs}.

As it is well known, one of the advantages of working with the
variable $v$ is that the quantum theory of primordial
perturbations is reduced to an action describing a free scalar
field with a time-dependent mass term. The question then is: which
are the appropriate observables that emerge from the quantum
theory encoded in the quantum field $\hat v$?

The standard answer is the power spectrum, which is normally
associated with the quantum two-point correlation function of the
quantum field $\hat v$. That is, the quantum theory of the
variable $v$, simultaneously sets the quantum theory of $\hat
\dphi$ and $\hat \Psi$ [see Eq. \eqref{defv}]. Afterwards, one
calculates the Fourier transform of $\bra 0 | \hat \Psi (\x,\eta)
\hat \Psi(\vec y,\eta) |0 \ket$ and relates it with the scalar
power spectrum of the curvature perturbation. In other words, in
the standard approach, one identifies the Fourier transform of the
quantum two-point correlation function with an average over an
ensemble of classical anisotropic universes of the same
correlation function:
\beq\label{identificacion} \bra 0 | \hat \Psi_{\nk} \hat
\Psi_{\nk'} |0 \ket = \overline{\Psi_{\nk} \Psi_{\nk'}} \equiv
2\pi^2 \delta(\nk + \nk') P_{\Psi} (k). \eeq

As mentioned in Sec. \ref{intro}, one usually encounters in the literature that
decoherence helps to understand the identification made in Eq.
\eqref{identificacion}, e.g. \cite{kiefer,jmartin}. The line of
reasoning is as follows: the dynamics of the inflationary universe
leads the vacuum state of the field $\hat v$ to a highly squeezed
state, and in this limit, all the quantum predictions can be
reproduced if one assumes that the system always followed
classical laws but had random initial conditions with a given
probability density function. Although we do not subscribe to
such posture (for a detailed analysis see Refs. \cite{shorts,stat}), that argument alone does not say
anything concerning the physical mechanism leading to a particular
realization of the field $\Psi$ corresponding to our universe. One
cannot apply the usual postulates of Quantum Mechanics based on
the Copenhagen interpretation, since entities such as
observers, measurements or measurement devices are
not well defined in the early universe. Moreover, even if in
principle there exist many universes, the fact is that we only
have observational access to one--our own--universe. Therefore,
the situation is completely different than the ordinary laboratory
setup, where one would check that the predictions provided by
the Quantum Theory can be verified by repeating the experiment
many times.

The previous described problem can be addressed by invoking a
self-induced collapse of the wave function
\cite{pss,shorts}. In particular, we assume that the vacuum
state associated to each mode of the field $\hat v_{\nk}$
spontaneously changes at a certain time $\tc$, called the time of
collapse, into a new state, i.e. $|0_{\nk} \ket  \to |
\Xi_{\nk} \ket$. The state $|\Xi_{\nk} \ket$ is no longer
invariant under rotations and spatial translations. Thus, the
post-collapse state characterizing the field is no longer
homogeneous and isotropic. These collapses for each mode will be
assumed to occur according certain rules called \emph{collapse
schemes}, and we will detail them in the next Section.

At this point, we must focus on the connection between the
classical and quantum prescriptions. In particular, here we will focus
on the scalar perturbation $\Psi$, representing the curvature
perturbation, which is intrinsically related to the temperature
anisotropies of the CMB. As precisely explained in Ref.
\cite{gabriel}, the relation between $\hat \Psi$ and $\Psi$
is made by taking the view that the classical description, encoded
in $\Psi$, is only relevant for those particular states for which
the quantity in question is sharply peaked and that the classical
description corresponds to the expectation value of said quantity.
For example, one can take the wave packet characterizing a free
particle, where clearly the wave function is sharply peaked around some
value of the position. In that context, one could claim that
the particle position is well defined and corresponds to the
expectation value of the position operator in that state described
by the wave packet. Given the previous discussion, we identify
\begin{equation}\label{psi}
\Psi (\x,\eta) = \bra \Xi |\hat  \Psi (\x,\eta) |\Xi \ket,
\end{equation}
with $|\Xi \rangle $ a state of the quantum field $\hat{v}(x)$
characterizing jointly the metric and the field perturbation, which
only acquires a physical meaning as long as the state
corresponds to a sharply peaked one associated to the quantum
field $\hat{\Psi}(x)$. In other words, after establishing the
quantum theory of $\hat v$, Eqs. \eqref{EEs} and \eqref{psi} imply
\beq\label{psi2}
\nabla^2 \Psi = \nabla^2 \bra \hat  \Psi   \ket =  -
\sqrt{\frac{\epsilon}{2}} \frac{H}{M_P} \left(\bra \hat v' \ket
-\frac{z'}{z} \bra \hat v \ket \right)
\eeq

It is worthwhile to mention that if we consider the vacuum state,
as it is in the standard approach, we would have $\bra 0 |\hat
\Psi (\x,\eta) | 0 \ket  = \Psi (\x,\eta) =0$. Consequently, the
space-time would be perfect homogeneous and isotropic. It is only
after the collapse that generically $\bra \Xi |\hat  \Psi
(\x,\eta) | \Xi \ket  = \Psi (\x,\eta) \neq 0$. This illustrates
how the metric perturbations are born from the self-induced
collapse.

After establishing how the primordial curvature perturbation is
generated within our approach, we can make contact with the
observational quantities. This is, we can extract the scalar power
spectrum from
\beq
\overline{\Psi_{\nk} \Psi_{\nk'} } = \overline{\bra \Xi_{\nk}
| \hat \Psi_{\nk} |  \Xi_{\nk} \ket \bra \Xi_{\nk'} | \hat
\Psi_{\nk'} |  \Xi_{\nk'} \ket} \label{bla}\eeq
The bar appearing in $\overline{\Psi_{\nk} (\eta) \Psi_{\nk'}
(\eta)}$ denotes an average over possible realizations of
$\Psi_{\nk}$, which is a random field and its randomness is
inherited by the stochastic nature of the collapse. In other
words, the average is over possible outcomes of the field
$\Psi_{\nk}$. The set of all modes of the field $\{\Psi_{\nk_1},
\Psi_{\nk_2}, \Psi_{\nk_3}, \ldots    \}$ characterizes a
particular universe $\mathcal{U}$. Thus, the average is over
possible realizations characterizing different universes
$\mathcal{U}_1, \mathcal{U}_2, \ldots$ Our universe, is just one
particular materialization $\mathcal{U}^*$. Note that this is
different form the standard inflationary account, in which the
power spectrum is obtained from $\bra 0 | \hat \Psi_{\nk} (\eta)
\hat \Psi_{\nk'} (\eta) | 0 \ket$, with all the mentioned
shortcomings. Meanwhile, in our picture, the power spectrum is
obtained from the expression $\overline{\bra \Xi_{\nk} | \hat
\Psi_{\nk} |  \Xi_{\nk} \ket \bra \Xi_{\nk'} | \hat \Psi_{\nk'} |
\Xi_{\nk'} \ket}$ where every element can be clearly justified.

Finally, the scalar power spectrum, within the collapse proposal,
is \cite{gabriel}:
\beq\label{PSscalar} P_{\Psi}(k) \propto \frac{H^2}{\epsilon M_P^2}
C(z_k) \eeq
with
\barr\nonumber
C(z_k) &\equiv&  \lambda_{\pi}^2
\left(1-\frac{1}{z_k^2}+\frac{1}{z_k^4}\right)\left[\cos
z_k-\frac{\sin z_k}{z_k}\right]^2 +\\&+&  \lambda_{v}^2
\left(1+\frac{1}{z_k^2}\right)\left[\frac{\cos
z_k}{z_k}-\left(\frac{1}{z_k^2}-1\right)\sin z_k\right]^2 \label{Ck}\earr
The parameters $\lambda_{\pi}$ and $\lambda_v$ can only take the
values 0 or 1 depending on which variable is affected by the
collapse, e.g. if only the momentum is affected by the collapse
then $\lambda_{\pi} = 1$ and $\lambda_v =0$. The parameter $z_k$
is defined as $z_k \equiv k \tc$, so it is directly related to
the time of collapse $\tc$. Therefore, the time of collapse
substantially modifies the scalar power spectrum in a very
particular manner that, in principle, can be used to distinguish
it from the traditional prediction. The specific technical details
regarding the implementation of the self-induced collapse
hypothesis that guided to result \eqref{PSscalar} can be consulted
in Refs. \cite{gabriel,alberto}; nevertheless, the steps are quite
similar to the ones that will be presented in the next section
concerning the tensor modes.

Another important aspect concerning the collapse scalar power
spectrum \eqref{PSscalar}, is that it is of the form $P_\Psi(k) =
\mathcal{A} C(z_k)$, which is different from the traditional
prediction $P_\Psi(k) = \mathcal{A} k^{n_s-1} $. The reason for this
apparent difference is because the collapse spectrum was obtained using the approximation that $\epsilon$ is exactly constant, thus, leading to $n_s = 1$. We could have worked with a better approximation in which $\epsilon'\neq 0 =$ constant, known as
quasi-de Sitter inflation and the final result would have been of
the form $P_\Psi (k)= \mathcal{A} Q(z_k) k^{n_s-1}$. However, it can be
shown \cite{lpl} that $Q(z_k) \simeq C(z_k)$ if the time of collapse occurs during the earlier stages of the inflationary regime; furthermore,
since in this article we are primarily interested in the amplitude
of the tensor modes, rather than the exact shape of their
spectrum, we can continue working in the approximation $\epsilon' =0$ and, thus, use the result \eqref{PSscalar}.

\section{Tensor modes and the tensor-to-scalar ratio}\label{tensor}

In order to proceed to find our results, in this Section we will study
the incorporation of the self-collapse hypothesis to the description of primordial tensor perturbations.

As it is known, these perturbations represent gravitational waves, and they are characterized by a symmetric, transverse and traceless tensor field. These properties lead to the existence of only two degrees of freedom. Therefore, the tensor $h_{ij}$, representing the gravitational waves, is usually decomposed as \cite{weinberg}:
\begin{equation}\label{cercana}
h_{ij}\,=\, h_+ e_{ij}^+ \, + \, h_{\times} e_{ij}^{\times}
\end{equation}
where $e_{ij}^{\alpha}$, $\alpha\,=+\,,\times$ is a time-independent polarization tensor. We will work with only one polarization $\alpha = +, \times$. As each polarization term is independent, and as each polarization leads to the same result, we will just multiply by a factor of two the spectrum associated to an individual case, at the end of our calculations, to obtain the final result.

The action for the gravitational waves can be obtained by expanding the Einstein action up to the second order in transverse, traceless metric perturbations $h_{ij}(\nx, \eta)$. The result is \cite{muk,mukbook},
\begin{equation}\label{accion1}
S\,=\,\frac{1}{64\pi G}\int d^3x  \, d\eta \, a^2 \left( h^{i\ '}_{\,j} h^{j\ '}_{\,i}  -  h^i_{\,j,l} h^{i\ ,l}_{\,j} \right)
\end{equation}
where the spatial indices are raised and lowered with the help of the unit tensor $\delta_{ik}$.

Then, we expand $h_{ij}$ in Fourier modes,
\begin{equation}\label{hacheij}
h_{ij}(\x, \eta) \, = \, \int \frac{d^3k}{(2\pi)^{3/2}} \, h_{\nk}(\eta) \, e_{ij}(\nk) \, e^{i\nk.\x},
\end{equation}
and substituting \eqref{hacheij} into the action \eqref{accion1} it is obtained,
\begin{equation}\label{accion2}
S=\frac{1}{64\pi G}\int d^3k  \, d\eta \, a^2 \, e^i_{\,j} e^j_{\,i} \, \left( h^{\ '}_{\nk} h^{\ '}_{-\nk} - k^2 h_{\nk} h_{-\nk} \right)
\end{equation}

Next, we perform the change of variable:
\begin{equation}
v_{\nk}=\sqrt{\frac{e^i_{\,j}e^j_{\,i}}{32\pi G}}\,a\,h_{\nk}
\end{equation}
and then, the action \eqref{accion2} can be rewritten as:
\begin{equation}\label{accion3}
S\,=\,\frac{1}{2} \int d^3k  \, d\eta \, \left[  \, v^{\,'}_{\nk} v^{\,'}_{-{\nk}} \, - \, \left( k^2 \,-\, \frac{a''}{a} \right) v_{\nk} v_{-{\nk}} \right].
\end{equation}
This action describes a real scalar field in terms of its Fourier transform,
\begin{equation}
v(\x, \eta)\, = \, \int \frac{d^3k}{(2\pi)^{3/2}} \, v_{\nk}(\eta) \, e^{i \nk.\x}
\end{equation}
Thus, the action for the variable $v(\x, \eta)$ results:
\begin{equation}\label{acc}
S=\frac{1}{2} \int d^3x  \, d\eta \, \left[ \, \left(v'\right)^2  -  \left( v_{,i} \right)^2  +  \frac{a''}{a} v^2  \right].
\end{equation}

Note that the momentum canonical to $v(\nx, \eta)$ is $\pi(\nx, \eta)\equiv \frac{\partial{(\sqrt{-g}\mathcal{L})}}{\partial v'} =v'(\nx, \eta)$.

In the quantization process, the field $v(\x, \eta)$ and its conjugate
momentum $\pi(\x, \eta)$ are promoted to operators acting on a Hilbert space
$\mathscr{H}$. These satisfy the standard equal time commutation
relations:
\begin{eqnarray}\nonumber
[\hv(\x, \eta), \hat{v}(\x', \eta)]&=&[\hp(\x, \eta), \hp(\x', \eta)]=0 \\
\nonumber [\hat{v}(\x, \eta), \hp(\x', \eta)]&=& i \delta (\x- \x')
\end{eqnarray}

The standard procedure is to decompose $\hat{v}$ and $\hat{\pi}$ in terms of the time-independent
creation and annihilation operators. For practical reasons, we
will work with periodic boundary conditions over a box of size
$L$, where $k_i L=2\pi n_i$ for $i=1,2,3$. So we write,
\begin{eqnarray}\nonumber
\hv(\nx, \eta) \, &=& \, \frac{1}{L^{3/2}} \, \sum_{\nk} \hv_{\nk}(\eta)e^{i \nk.\nx}\\
\nonumber \hp(\nx, \eta) \, &=& \, \frac{1}{L^{3/2}} \, \sum_{\nk} \hp_{\nk}(\eta)e^{i \nk.\nx}
\end{eqnarray}
where $\hv_{\nk}(\eta)= v_k(\eta) \hb_{\nk}+v^*_k(\eta) \hb^\dagger_{-\nk}$ and $\hp_{\nk}(\eta)=v'_k(\eta) \hb_{\nk}+v'^*_k(\eta) \hb^\dagger_{-\nk}$.

The mode functions $v_k(\eta)$ are normalized such that
\begin{equation}\label{norm}
v^*_k v'_k \, - \, v_k v'^*_k \, = \, -i,
\end{equation}
and then, the creation and annihilation operators $\hb_{\nk}$ y $\hb^\dagger_{\nk}$ satisfy the commutation relations:
\begin{eqnarray}\nonumber
[\hb_{\nk}, \hb_{\nk'}] &=& [\hb^\dagger_{\nk}, \hb^\dagger_{\nk'}] = 0 \\
\nonumber [\hb_{\nk}, \hb^\dagger_{\nk'}] &=&  \delta (\nk-\nk')
\end{eqnarray}

From \eqref{acc}, the equation of motion for $v_{\nk}$ results:
\begin{equation}
v''_k \, + \, \left( k^2 \, - \, \frac{a''}{a} \right) v_k \, = \, 0
\end{equation}
As it was mentioned earlier, we are working in an approximately \textit{de Sitter} inflation where $H \simeq \rm{const.}$, and hence $a(\eta) \simeq -1/(H\eta)$. Using this approximation, the last equation takes the form:
\begin{equation}
v''_k \, + \, \left( k^2 \, - \, \frac{2}{\eta^2} \right) v_k \, = \, 0
\end{equation}
whose solution, choosing the Bunch-Davies vacuum as the initial state, is:
\begin{equation}\label{vsubk}
v_k \, = \, \frac{1}{\sqrt{2k}}  \left(1\,-\, \frac{i}{k\eta} \right) e^{-ik\eta}.
\end{equation}

At this point, we introduce the self-induced collapse proposal: we suppose that at time (dependent on the mode $k$), $\eta_k^c$, called the \textit{time of collapse}, the vacuum state associated to each mode of the field $\hat v_{\nk}$ spontaneously changes into a new state, i.e. $|0_{\nk} \ket  \to |
\Xi_{\nk} \ket$. The state $|\Xi_{\nk} \ket$ is no longer invariant under rotations and spatial translations. Thus, the
post-collapse state characterizing the field is no longer homogeneous and isotropic. We will not consider a specific
collapse mechanism, but we will parameterize the collapse through the expectation values of the field and its conjugated momentum
evaluated in the post-collapse state, as it will be shown below.

In order to proceed, we decompose the operators $\hv_{\nk}(\eta)$ and $\hp_{\nk}(\eta)$ in their real and imaginary parts,
\begin{eqnarray}\label{lejana}
\hv_{\nk}(\eta) \, &=& \, \hv^R_{\nk}(\eta) \, + \, i \hv^I_{\nk}(\eta)\\
\hp_{\nk}(\eta) \, &=& \, \hp^R_{\nk}(\eta) \, + \, i \hp^I_{\nk}(\eta)
\end{eqnarray}
where
\begin{equation}\label{uso1}
\hv^{R,I}_{\nk} (\eta) \, = \, \frac{1}{\sqrt{2}} \bigg( v_k(\eta) \hb^{R,I}_{\nk} \, + \, v^*_k(\eta) \hb^{\dagger\,R,I}_{\nk} \bigg),
\end{equation}
and
\begin{equation}\label{uso2}
\hp^{R,I}_{\nk} (\eta) \, = \, \frac{1}{\sqrt{2}} \bigg( v'_k(\eta) \hb^{R,I}_{\nk} \, + \, v'^*_k(\eta) \hb^{\dagger\,R,I}_{\nk} \bigg)
\end{equation}
with $\hb^R_{\nk}=\frac{1}{\sqrt{2}} (\hb_{\nk}+\hb_{-\nk})$ and $\hb^I_{\nk}=\frac{-i}{\sqrt{2}} (\hb_{\nk} - \hb_{-\nk})$.
In this manner, $\hat{v}_{\nk}^{{R,I}}(\eta)$ and $\hat{\pi}_{\nk}^{{R,I}}(\eta)$ are Hermitian operators, which we know from standard Quantum Mechanics that these kind of operators can be subjected to a ``measurement'' type of process.

The commutation relations for these operators read,
\begin{eqnarray}\nonumber
[\hb^R_{\nk}, \hb^{\dagger\, R}_{\nk'}]&=&\delta_{\nk,\nk'}\,+\,\delta_{\nk,-\nk'}\\ \label{conmdep}
[\hb^I_{\nk}, \hb^{\dagger\, I}_{\nk'}]&=&\delta_{\nk,\nk'}\,-\,\delta_{\nk,-\nk'}
\end{eqnarray}
with all the other commutators vanishing. Note that, in the last equation, $\nk$ and $-\nk$ are not independent.

Being $\hv^{R,I}_{\nk} (\eta)$ and $\hp^{R,I}_{\nk} (\eta)$ Hermitian operators, we will
evaluate their expectation values:
\begin{eqnarray}\nonumber
\langle \hv^{R,I}_{\nk} (\eta^c_k) \rangle_{\Xi}&=& \lambda_v x^{R,I}_{\nk,1}\sqrt{\Big[\Delta\hv(\eta^c_k)\Big]^2_0}\\
&=&\lambda_v x^{R,I}_{\nk, 1} \frac{1}{\sqrt{2}} |v_k(\eta_k^c)| \label{colaps1}
\end{eqnarray}
\begin{eqnarray}\nonumber
\langle \hp^{R,I}_{\nk} (\eta^c_k) \rangle_{\Xi}&=& \lambda_{\pi} x^{R,I}_{\nk,2}\sqrt{\Big[\Delta\hp(\eta^c_k)\Big]^2_0}\\
&=&\lambda_{\pi} x^{R,I}_{\nk, 2} \frac{1}{\sqrt{2}} |v'_k(\eta_k^c)| \label{colaps2}
\end{eqnarray}
where, the numbers $x^{R,I}_{\nk, 1}$ and $x^{R,I}_{\nk, 2}$ are a collection of independent random quantities (selected from a Gaussian distribution centered at zero with unit-spread), and $[\Delta\hv(\eta^c_k)]^2_0$ and $[\Delta\hp(\eta^c_k)]^2_0$ are the quantum uncertainties of the operators $\hat{v}_{\nk}^{{R,I}}$ and $\hat{\pi}_{\nk}^{{R,I}}$ in the vacuum state $|0\rangle$ at time $\eta^{c}_{k}$.

The parameters $\lambda_v$ and $\lambda_\pi$ are viewed as ``switch-off/on''
parameters. This is, they can only take the values 0 or 1 depending on which variable
$\hv^{R,I}_{\nk}$, $\hp^{R,I}_{\nk}$ or both is affected by the collapse. For instance,
in past works \cite{pss,multiples}, the name \emph{independent
scheme} was coined for the case $\lambda_v=\lambda_\pi=1$, i.e.  $\hv^{R,I}_{\nk}$ and
$\hp^{R,I}_{\nk}$ are both affected independently by the collapse. Nevertheless, there
are other options, e.g. $\lambda_v=0$ and $\lambda_\pi=1$. In the rest of the present
Letter, we will keep the $\lambda_v$ and $\lambda_\pi$ parameters without referring to a
particular collapse scheme.

Given a post-collapse state $|\Xi\rangle$, next to equations (\ref{uso1}) and (\ref{uso2}), it can be seen that,
\begin{equation}\label{real1}
\langle \hv^{R,I}_{\nk} (\eta) \rangle_\Xi = \sqrt{2} \, \Re\: \Big[v_k(\eta) \langle \hb^{R,I}_{\nk} \rangle_\Xi \Big]
\end{equation}
\begin{equation}\label{real2}
\langle \hp^{R,I}_{\nk} (\eta) \rangle_\Xi = \sqrt{2} \, \Re\: \Big[v'_k(\eta) \langle \hb^{R,I}_{\nk} \rangle_\Xi \Big]
\end{equation}

Now, we evaluate (\ref{real1}) and (\ref{real2}) at time of collapse $\eta_k^c$. This allows us to obtain an expression for $\langle \hb^{R,I}_{\nk} \rangle_\Xi$ in terms of the quantities $\langle \hv^{R,I}_{\nk} (\eta_k^c) \rangle_\Xi$ and $\langle \hp^{R,I}_{\nk} (\eta^c_k) \rangle_\Xi$. Once this is done, we can rewrite (\ref{real1}) which now reads:
\begin{eqnarray}\nonumber
&&\langle \hv^{R,I}_{\nk} (\eta) \rangle_\Xi = \langle \hv^{R,I}_{\nk} (\eta_k^c) \rangle_\Xi \Bigg\{ \left( 1+\frac{1}{k\eta z_k}-\frac{1}{z_k^2} \right) \times\\
\nonumber &\times& \cos(k\eta-z_k)+\left[ \frac{1}{k\eta} \left( \frac{1}{z_k^2}-1 \right)+\frac{1}{z_k} \right] \sin(k\eta-z_k) \Bigg\} +\\
\nonumber &+&\frac{\langle \hp^{R,I}_{\nk} (\eta_k^c) \rangle_\Xi}{k} \Bigg\{ \left(\frac{1}{k\eta}-\frac{1}{z_k} \right) \cos(k\eta-z_k) +\\
&+&\left( 1+\frac{1}{k\eta z_k} \right) \sin(k\eta-z_k) \Bigg\}\label{larga}
\end{eqnarray}
where $z_k \equiv k \eta_k^c$.

Working with equations (\ref{vsubk}), (\ref{colaps1}), (\ref{colaps2}) and (\ref{larga}), we arrive to the expression:
\begin{equation}\label{vri}
\langle \hv^{R,I}_{\nk} (\eta) \rangle_\Xi = \frac{1}{2k^{1/2}} \left[ \lambda_v
F(k\eta, z_k) x^{R,I}_{{\nk}, 1} + \lambda_\pi G (k\eta, z_k) x^{R,I}_{{\nk}, 2} \right]
\end{equation}
where,
\begin{eqnarray}\nonumber
&&F(k\eta, z_k)\equiv  \left(1\,+\,\frac{1}{z_k^2} \right)^{1/2} \Bigg\{
\left(1+\frac{1}{k\eta z_k}-\frac{1}{z_k^2} \right)\times\\
\nonumber &&\times \cos(k\eta-z_k)+ \left[ \frac{1}{k\eta} \left( \frac{1}{z_k^2}-1 \right)+\frac{1}{z_k} \right] \sin(k\eta-z_k)  \Bigg\}
\end{eqnarray}
\begin{eqnarray}\nonumber
&&G(k\eta, z_k) \equiv \left[\frac{1}{z_k^2}+ \left( 1- \frac{1}{z_k^2}
\right)^2 \right]^{1/2} \Bigg\{ \left(\frac{1}{k\eta}-\frac{1}{z_k} \right)\times\\
\nonumber &&\times \cos(k\eta-z_k)+ \left( 1+\frac{1}{k\eta z_k} \right) \sin(k\eta-z_k)
\Bigg\}
\end{eqnarray}

On the other hand, from equation (\ref{lejana}), we will evaluate the expectation value of $\hv_{\nk}(\eta)$ in the post-collapse state,
\begin{equation}
\langle \hv_{\nk}(\eta) \rangle_\Xi =\langle \hv^R_{\nk}(\eta) \rangle_\Xi + i \langle \hv^I_{\nk}(\eta) \rangle_\Xi
\end{equation}
By using (\ref{vri}), we obtain:
\begin{equation}
\langle \hv_{\nk}(\eta) \rangle_\Xi = \frac{1}{2k^{1/2}} \left[\lambda_v F(k\eta, z_k)
x_{{\nk},1} +\lambda_\pi G (k\eta, z_k) x_{{\nk}, 2} \right]
\end{equation}
where $x_{{\nk}, j} = x^R_{{\nk}, j}\,+\,i x^I_{{\nk}, j}$ with $j=1, 2$.

Since we have quantized $\hv_{\nk}(\eta)$, we can return to the original variable $h_{ij}(\x, \eta)$, describing the metric tensor perturbations. Therefore, we find that,
\begin{equation}
\hh_{ij}(\x, \eta) = \frac{1}{L^{3/2}} \sum_{\nk} \hh_{\nk}(\eta) e_{ij}({\nk}) e^{i {\nk}.\x}
\end{equation}
where
\begin{equation}
\hh_{\nk}(\eta) =  \sqrt{\frac{32 \pi G}{e^i_{\,j} e^j_{\,i}}} \frac{1}{a(\eta)} \hv_{\nk}(\eta)
\end{equation}

Evaluating the expectation value of the last quantity, in the post-collapse state, we obtain:
\begin{equation}\label{igualdadchingona}
\langle \hh_{\nk}(\eta) \rangle_\Xi = \sqrt{\frac{32 \pi G}{e^i_{\,j} e^j_{\,i}}} \frac{1}{a(\eta)} \langle \hv_{\nk}(\eta) \rangle_\Xi
\end{equation}

Similarly to what was said to the equation \eqref{psi}, here we will identify
\begin{equation}\label{ident}
\langle \hh_{\nk}(\eta) \rangle_\Xi \simeq h_{\nk} (\eta)
\end{equation}
This means that the expectation value of $\hh_{\nk}$ coincides approximately with the
amplitude value of the classical gravitational wave $h_{\nk}$. After this identification
is made, we can evaluate the classical amplitude during the inflationary phase. Since we
are considering \textit{slow-roll} inflation, and because we are working in the
approximation $a(\eta) \simeq -1 / H \eta$, the classical amplitude results:
\begin{eqnarray}
h_{\nk}(\eta)&=&\frac{2H}{{M_P}}\frac{(-\eta)}{\sqrt{e^i_{\,j}e^j_{\,i}}}\frac{1}{k^{1/2}}
\Bigg[ \lambda_v F(k\eta, z_k) x_{{\nk},1} \nonumber \\
&+& \lambda_\pi G (k\eta, z_k) x_{{\nk}, 2} \Bigg]
\end{eqnarray}

As it is usual in the literature, if the Hubble radius is representative of the horizon,
the observational relevant modes are those satisfying the condition $k \ll \h$. Since
during inflation $\mH \simeq  -1/\eta$, the condition for modes that are outside the
horizon becomes $-k\eta \to 0$. In this limit, it can be shown that:
\begin{eqnarray}\nonumber
&&\lim_{-k\eta \,\to\, 0} F(k\eta, z_k) \, = \, \left( \frac{1}{-k\eta} \right) f(z_k) \\
\nonumber &&\lim_{-k\eta \,\to\, 0} G(k\eta, z_k) \, = \, \left( \frac{1}{-k\eta} \right)
g(z_k)
\end{eqnarray}
where,
\begin{equation}\label{fexacto}
f(z_k) \equiv \left(1+\frac{1}{z_k^2} \right)^{1/2} \left[-\frac{1}{z_k}
\cos(z_k)+\left(\frac{1}{z_k^2}-1\right) \sin(z_k)\right]
\end{equation}
\begin{equation}\label{gexacto}
g(z_k) \equiv  \left(1-\frac{1}{z_k^2}+\frac{1}{z_k^4}\right)^{1/2}
\left(-\cos(z_k)+\frac{1}{z_k} \sin(z_k)\right)
\end{equation}

Therefore, for modes outside the horizon we obtain:
\begin{equation}\label{chin}
h_{\nk}(\eta)  =  \frac{2H}{{M_P}}  \frac{1}{\sqrt{e^i_{\,j}e^j_{\,i}\,}}
\frac{1}{k^{3/2}}  \Bigg[\lambda_v f(z_k) x_{{\nk},1}  + \lambda_\pi g (z_k) x_{{\nk}, 2}
\Bigg]
\end{equation}
This quantity is approximately constant (since $H  \simeq$ const.). Additionally, it
depends on the random numbers $x_{{\nk},1}$ and $x_{{\nk},2}$, and also on the time of
collapse through the variable $z_k \equiv k\eta_k^c$. Note that this expression is only
possible by considering the self-induced collapse, and every element has a clear physical
origin. It has no counterpart in the traditional approach, where $h_k (\eta)$ is only
assumed to acquire a classical meaning somehow (e.g. decoherence, squeezing of the vacuum
state, many-world interpretation of Quantum Mechanics, etc.) only after the proper
wavelength associated to the mode $k$ becomes bigger than the Hubble radius $H^{-1}$.

Now, considering that $x_{{\nk}, 1}^{R,I}$ and $x_{{\nk}, 2}^{R,I}$ are independent random
numbers, and since ${\nk}$ and $-{\nk}$ are not independent quantities, we have:
\begin{equation}
\overline{x_{{\nk}, i}^R x_{{\nk}', i}^R} = \delta_{{\nk},{\nk}'} + \delta_{{\nk},-{\nk}'}
\end{equation}
\begin{equation}
\overline{x_{{\nk}, i}^I x_{{\nk}', i}^I} = \delta_{{\nk},{\nk}'} - \delta_{{\nk},-{\nk}'}
\end{equation}
where $i = 1, 2$. This leads to:
\begin{equation}
\overline{x_{{\nk}, 1} x^*_{{\nk}', 1}} = \overline{x_{{\nk}, 2} x^*_{{\nk}', 2}} =
2\delta_{{\nk},{\nk}'}
\end{equation}
and because $x_{{\nk}, 1}$  and $x_{{\nk}', 2}$ are not correlated,
\begin{equation}
\overline{x_{{\nk}, 1} x^*_{{\nk}', 2}} = 0
\end{equation}

Thus, from equation (\ref{chin}) we arrive to:
\begin{equation}
\overline{h_{\nk}(\eta) h_{{\nk}'}^{*}(\eta)} = \frac{8H^2}{M_P}
\frac{1}{e^i_{\,j}e^j_{\,i}} \frac{1}{k^3} \Bigg[ \lambda_v^2 f^2(z_k) + \lambda_\pi^2
g^2(z_k) \Bigg]\, \delta_{{\nk},{\nk}'}\label{bla2}
\end{equation}

As discussed previously for \eqref{bla}, from \eqref{bla2} the power spectrum for the
primordial gravitational wave amplitudes can be extracted. We obtain:
\begin{equation}\label{especttensorial}
P_h(\eta) = \frac{H^2}{\pi^5 M_P^2}\, C(z_k)
\end{equation}
where $C(z_k)\equiv \lambda_v^2 f^2(z_k)+\lambda_\pi^2 g^2(z_k)$ and it coincides exactly
with \eqref{Ck}.

Since any dependence on $k$ is in the function $C(z_k)$ through $z_k \equiv k \eta_k^c$,
if the time of collapse scales as $\eta_k^c \propto 1/k$, then $z_k$ is independent of
$k$. In this manner, the power spectrum \eqref{especttensorial} [as in the scalar case
\eqref{PSscalar}] becomes a scale free spectrum. Also, small variations in the relation
$\eta_k^c \propto 1/k$ would yield deviations in the spectrum shape with respect to the
standard prediction, which could be observationally distinguished.

Finally, from equations \eqref{especttensorial} and \eqref{PSscalar}, we can evaluate the
tensor-to-scalar ratio $r$. This quantity results to be:
\begin{equation}\label{ratiodef}
r \, \equiv \, \frac{P_h}{P_\Psi} \, \propto \, \frac{(H^2/M_P^2)\, C(z_k)}{(H^2/M^2_P \,
\epsilon) C(z_k)}
\end{equation}
Hence, this means:
\begin{equation}\label{rfinal}
r \, \propto \, \epsilon
\end{equation}

A few remarks are in order. According to latest observations from Planck mission, the
scalar power spectrum is practically scale invariant \cite{planck1}; on the other hand,
even if the tensor power spectrum is also expected to be close to scale invariant, the
fact is that detection of primordial gravity waves is still waiting for confirmation
\cite{bicep, planck2}.

As is clear from expression \eqref{rfinal}, the prediction for the tensor-to-scalar ratio
$r$ is independent of our model parameters. In particular, it does not depend on the time
of collapse. This can be seen from Eqs. \eqref{Ck} and \eqref{especttensorial} where the
modification to both power spectra
(scalar and tensor) is given by exactly the same function $C(z_k)$. Therefore, a possible
confirmation regarding the detection of primordial gravitational waves will not help to
constraint the collapse parameters, but only will set, as in the standard case, an energy
scale for inflation. The constriction of the collapse parameters can be made by focusing
on the scalar power spectrum and also the primordial bispectrum \cite{bispec}.

The fact that $r$ is independent of the collapse parameters can be understood as
follows. The quantum theory of the scalar perturbations, using the Mukhanov-Sasaki
variable, can be considered as a theory representing a collection of parametric
oscillators (i.e. one oscillator per mode), whose time-dependent frequency can be
expressed as $\omega^2_s (\eta,k) = k^2 -z''/z$. Furthermore, the quantum theory of the
tensor perturbations can also be viewed as a theory representing a collection of
parametric oscillators [see Eq. \eqref{accion3}]. In this case, the time-dependent
frequency is given by $\omega^2_t (\eta,k) = k^2 -a''/a$. Additionally, $z''/z =
a''/a$ up to first-order in the slow-roll parameters. Therefore, the physical mechanism
behind
what we effectively describe as a self-induced collapse, should not in principle
distinguish between the quantum theory of the scalar and tensor perturbations because
they are essentially the same, i.e. a collection of harmonic oscillators with a
time-dependent frequency that happens to be practically the same in both cases.
As matter of fact, the parameters $\lambda_v$ and $\lambda_\pi$ that control which
variable is affected by the collapse (recall that the values of these parameters can
only be 0 or 1 depending on which field $\hat v_{\nk}$ or $\hat
\pi_{\nk}$ or both is affected by the collapse) should be the same for the scalar and
tensor modes because there is no difference in the quantum theory characterizing the
scalar and tensor perturbations. We
think this is the main reason behind the fact that the modification to both power spectra
is given by the same function $C(z_k)$ and consequently $r$ is independent of the
collapse parameters.

Moreover, as it was mentioned in Sec. \ref{review}, in order for our
model prediction for the scalar
power spectrum to be consistent with CMB data, the time of collapse must satisfy $\tc
\propto 1/k$. That is, if the tensor power spectrum is also expected to be close to
scale invariant, then the time of collapse must also be of the form $\tc \propto 1/k$.
Thus, the dependence on the wave number $k$ of the time of collapse is exactly the same
for the scalar and tensor modes. This result is consistent with our previous discussion
in the sense that the self-induced collapse somehow affects all kind of perturbations
(scalar and/or tensor) in the same way. On the contrary,  the situation in which the
self-induced collapse
proposal is based on the semiclassical gravity framework, is different from the one based
on the Mukhanov-Sasaki variable. That is, in the semiclassical gravity approximation the
source terms that generate the curvature perturbations does not affect equally the scalar
and tensor modes, consequently in that approach the prediction for $r$ is different as in
the present work \cite{secorder}.

It is worthwhile to mention that in the expression for the scalar power spectrum, Eq.
\eqref{PSscalar}, the slow-roll parameter $\epsilon$ appears explicitly, while in the
expression for the tensor power spectrum Eq. \eqref{especttensorial} it does not. The
reason for this difference can be traced back in the way we have linked the scalar and
tensor curvature perturbations to the quantum variables affected by the collapse. The
scalar curvature perturbation $\Psi_{\nk}$ is generated by evaluating the field variables
$\hat v_{\nk}$ and $\hat v_{\nk}'$ (which is essentially $\hat \pi_{\nk}$) at the
post-collapse state, Eq. \eqref{psi2}. In this expression, $\epsilon$ appears explicitly
and it was obtained using Einstein equations. On the other hand, the tensor curvature
perturbation is generated by the expectation value of $\hat v_{\nk}$ only, Eq.
\eqref{igualdadchingona}, which is independent of the slow-roll parameter.\footnote{Note
that in Eq. \eqref{psi2} the slow-roll parameter
$\epsilon$ appears in the numerator, while in the expression for the scalar
power spectrum \eqref{PSscalar} appears in the denominator. The reason for
this difference is that, in the longitudinal gauge, the scalar curvature
perturbation $\Psi$ becomes amplified by a factor of $1/\epsilon$ during the
transition from inflation to the radiation dominated stage
\cite{gabriel2010,deruelle}, in which the
CMB is originated. Consequently, in order to obtain a consistent
prediction to be compared with the observations, we must multiply by a
factor of $1/\epsilon^2$ the scalar power spectrum obtained during inflation
associated to $\overline{\Psi_{\nk} \Psi_{\nk'}}$.}

Furthermore, the fact that $r \propto \epsilon$, within the framework of the present
manuscript, makes this prediction indistinguishable from the standard case. However, this
only applies to the amplitude of the tensor modes. The scalar power spectrum is
substantially different from the traditional inflationary paradigm. The difference is
encoded in the function $C(z_k)$, and one can perform an analysis using the observational
data, as the one done in e.g. \cite{susana2012}. Additionally, a possible improvement in
future experiments, regarding the detection of the shape and amplitude of the primordial
bispectrum, can also help to discriminate between our proposal and the standard prediction
\cite{bispec}. Moreover, the main consequence of the result obtained in this Letter is
that a confirmed detection of a non-vanishing value for $r$ can differentiate between the
two frameworks of the self-induced collapse proposal, namely, the semiclassical gravity
approach and the joint matter-metric quantization, as reflected in the quantization of the
Mukhanov-Sasaki variable. In the former case, the predicted value for $r$ is suppressed
by a factor of $10^{-9} \epsilon^2$ \cite{secorder}; thus, practically undetectable.
While in the latter, $r$ is of the same
order of magnitude as the slow-roll parameter $\epsilon$ and, hence,
from the observational point of view, in the same footing as the standard picture.

\section{Conclusions}\label{conclusions}

As it has been mentioned in previous works e.g. \cite{pss,gabriel2010}, working in the
framework of semiclassical gravity, the collapse hypothesis, which serves to address the
transition from an homogeneous and isotropic state to another one which is not, leads to a
practically undetectable amplitude for the primordial gravitational waves. For this
reason, and motivated by the implications of a possible detection of primordial $B$
polarization modes, we have  calculated the amplitude of tensor modes in the joint
metric-matter quantization of the primordial perturbations, but taking into account the
self-induced collapse hypothesis. We have accomplished this task by assuming a slow-roll
type of
inflation and characterizing the collapse by the expectation values of the field and its
conjugated momentum; in this sense, we have considered a generic type of collapse.

It is also worthwhile to mention that our approach differs drastically from the one considered in Ref. \cite{hinduesT}. As mentioned in the Introduction, our point of view is that the quantum collapse is directly related to the generation of the primordial perturbations. Therefore, if there is no quantum collapse, then $\Psi_{\nk} = 0=h_{\nk}$. In turn, the authors in Ref. \cite{hinduesT} consider a particular collapse mechanism, known as CSL, and apply it directly to the Mukhanov-Sasaki variable obtaining a prediction for $r$ (as well as for the scalar and tensor power spectra) that depends on the CSL model parameters. However, in their work, if there is no quantum collapse, then $P_{\Psi}$, $P_h$ and $r$ are exactly the same as in the standard approach, thus, changing drastically the physical implication of assuming a self-induced collapse, as well as, the theoretical prediction for $r$.

Our results indicate that it is possible to obtain a detectable amplitude associated to
the primordial gravitational waves even by adding the self-induced collapse hypothesis.
The predicted amplitude is quite similar to the one provided by standard inflation, i.e.
$r \propto \epsilon$. This result implies that our model prediction is consistent with
the latest findings from the joint BICEP/Planck collaboration \cite{bicepplanck}. Also,
as a consequence of our result $r \propto \epsilon$, the amplitude is independent of the
collapse mechanism; particularly, is independent of the time of collapse $\tc$. Therefore,
even if the power spectra (scalar and tensor) do depend on $\tc$ each one, they do in the
exactly same way making $r$ independent of the time of collapse. On the other hand, a
detection of primordial gravity waves cannot help to distinguish between the collapse
proposal \emph{\`{a} la} Mukhanov-Sasaki and the standard inflation case. In order to
discriminate between the two approaches, one must focus on the scalar power spectrum and
the primordial bispectrum.

Finally, if a detection of primordial gravitational waves is confirmed, and consequently,
$r$ turns out to be non-vanishing, the collapse hypothesis applied to the inflationary
universe, within the framework of the semiclassical gravity approximation, would face
serious issues; in consequence, the most viable option would be to consider the
self-induced collapse applied to the Mukhanov-Sasaki variable.  Therefore, the result
obtained in this work, along with future observational data, can help to improve our
overall knowledge of the collapse mechanism behind the primordial perturbations; in
particular, the relation between the collapse and the gravitational aspects in the early
universe.

\bigskip

\acknowledgments{G.R.B. is supported by CONICET (Argentina).
G.R.B. acknowledge support from the PIP 112-2012-0100540 of
CONICET (Argentina).  G.L.'s research
funded by Consejo Nacional de Investigaciones Cient\'{i}ficas y T\'{e}cnicas,
CONICET (Argentina) and by Consejo Nacional de Ciencia y Tecnolog\'{i}a, CONACYT
(Mexico).}

\end{document}